# Photon Counting EMCCDs: New Opportunities for High Time Resolution Astrophysics

Craig Mackay*[a], Keith Weller[a], Frank Suess[a]

[a]Institute of Astronomy, University of Cambridge, Madingley Road, Cambridge CB3 0HA, UK

**ABSTRACT**

Electron Multiplying CCDs (EMCCDs) are used much less often than they might be because of the challenges they offer camera designers more comfortable with the design of slow-scan detector systems. However they offer an entirely new range of opportunities in astrophysical instrumentation. This paper will show some of the exciting new results obtained with these remarkable devices and talk about their potential in other areas of astrophysical application. We will then describe how they may be operated to give the very best performance at the lowest possible light levels. We will show that clock induced charge may be reduced to negligible levels and that, with care, devices may be clocked at significantly higher speeds than usually achieved. As an example of the advantages offered by these detectors we will show how a multi-detector EMCCD curvature wavefront sensor will revolutionise the sensitivity of adaptive optics instruments and been able to deliver the highest resolution images ever taken in the visible or the near infrared.

## 1. INTRODUCTION

Most astronomical objects are large and, since light variations cannot occur on timescales shorter than the light travel time across an object, we expect most objects to vary only very slowly in brightness. Recently, however, it has become clear that some important astronomical objects are actually very small and that fluctuations may be detected on very short timescales indeed. For example, stellar flares and pulsations can occur in a fraction of a second and the surface oscillations of compact stars such as white dwarfs and neutron stars can occur on microsecond timescales. There is increasing interest in extreme binaries and accretion disc objects where millisecond timescales are often encountered. A great deal can be learned of the physics of objects by comparing the timing of outbursts in the visible and at x-ray wavelengths simultaneously. Although pulsars emit very little energy in the visible part of the spectrum, the prospect of being able to observe them with millisecond resolution on the next-generation of optical telescopes is very exciting.

In addition to the wish to observe objects that are themselves varying rapidly there is the additional challenge we have with ground-based observatories. It is that atmospheric fluctuations induce changes in the quality of an astronomical image on timescales a short as a few milliseconds. Techniques for overcoming these fluctuations such as adaptive optics require high-speed detector systems that detect and deliver significant data about the fluctuations on short timescales.

Historically high-speed astronomical detectors working in the visible part of the spectrum have been fairly limited in their capabilities. Photomultipliers and avalanche photodiodes are generally single element detectors or arrays of only a few pixels. Imaging detectors such as intensified photon counting detectors have other characteristics that makes their application to astronomical imaging and spectroscopy less attractive. Increasingly, however, instrument builders are turning to charge coupled devices to provide the detection capabilities at very high-speed. Although for direct imaging and spectroscopy CCDs are now ubiquitous as detectors, when operated at high speed they are limited by the noise of the on chip buffer amplifier through which all the output signals pass. While these can operate at readout noise levels of only a few electrons at pixel rates in the tens-hundreds of kilohertz range, at high speed (tens of megahertz) the read noise can become prohibitive, particularly when imaging relatively faint astronomical targets.The recent development of electron multiplying CCDs by E2V Technologies (Chelmsford, UK) Ltd (figure 1) has revolutionised high-speed widefield imaging.

*cdm <at> ast.cam.ac.uk

Similar devices were made by Texas Instruments but unfortunately TI have recently withdrawn their devices from production and so we astronomers are now left with a single manufacturer. EMCCDs have internal gain and are able to run at high-speed. They allow high efficiency photon counting operation even at the highest speed. This paper will discuss some of the capabilities and limitations of these devices as they affect astronomy. As an example, we will look at how they have revolutionised high angular resolution ground-based astronomy through lucky imaging and what impact EMCCDs are likely to have on the performance of ground-based telescopes over the next few years.

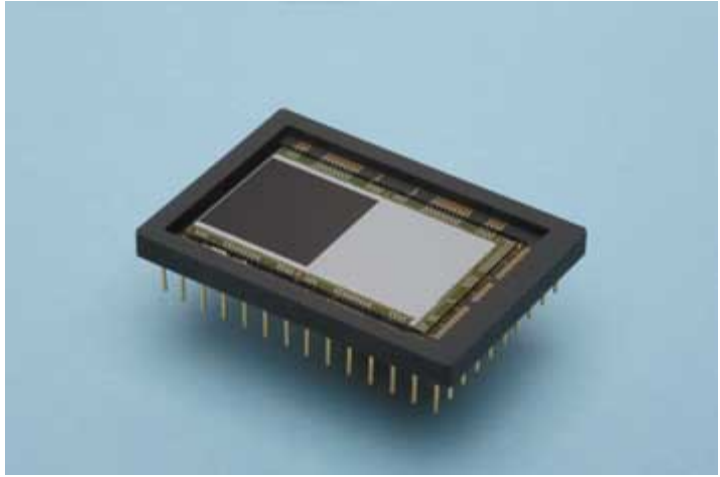

Figure 1: The CCD 201 manufactured by E2V technologies Ltd (Chelmsford, UK). This is a back illuminated (thinned) electron multiplying CCD with 1024 x 1024 pixels of 13 mu.²

## 2. THE ARCHITECTURE OF ELECTRON MULTIPLYING CCDS

Almost all the properties of electron multiplying CCDs are indistinguishable from those of conventional charge coupled devices. The charge generation, storage and charge transfer mechanisms are the same allowing integrations over long periods of time with high quantum efficiency (particularly in the case of back illuminated/thinned CCDs). This is very important because the extraordinary progress that has been made in recent years by manufacturers such as E2V Technologies in delivering extremely high quality devices in terms of device uniformity, high quantum efficiency and cosmetic quality directly benefits the development of EMCCDs. The only difference between a conventional CCD and electron multiplying CCD is the extension of the output register. A multiplication output register is placed between the normal serial transfer output register and the output amplifier. Within this multiplication output register one of the electrodes is clocked with a much higher voltage swing (30-40 V swing) than the ~12 volt swing typically used for both parallel (image) and serial registers. When an electron is transferred through one of these gates there is a small probability ( ~1-1.5 %) that an avalanche multiplication occurs so that where there was one electron before there are now two. This process continues as the charge moves down the multiplication register. The output signal is then on average 1.01 or 1.015 to a high power where that power is the number of elements in the multiplication register. The CCD201 from E2V Technologies Ltd has 536 elements in the multiplication register giving a gain of about 205 or 2920 respectively. Beyond the multiplication register the readout amplifier is a normal high-speed amplifier, but the effective readout noise of that amplifier is now reduced by this multiplication factor. With a typical unity-gain read noise at 30 MHz of about 100 electrons it is clear that with a relatively high gain level an individual photon may be detected with a very high signal-to-noise. The gain level may be adjusted by changing the high clock voltage level.

Accumulating large numbers of images taken at very low signal level (much less than one photon per pixel per frame) produces images which are perfectly sharp and perfectly aligned with an image taken at a much higher signal level[1]. It is extraordinary that individual electrons can be transported over distances of up to 2 cm without getting out of position by as much as one 13 µ pixel. The characteristics of EMCCDs are very similar to those of standard CCDs in terms of high quantum efficiency, charge transfer efficiency and cosmetic quality. Inevitably, working at very high gains allows us to see aspects of the way that CCDs work which are normally hidden from us with conventional slow scan operated CCDs. There are two features of EMCCDs which need to be considered carefully when designing a camera system for a particular application. The first is the effect of the stochastic multiplication process that adds an additional variance to the output signal. The second is broadly known as clock induced charge (CIC).

The EMCCD produces an output level on average per photon equal to the gain. However, the stochastic nature of the multiplication process firstly makes the most probable gain to be unity, giving an output level of one electron. The probability distribution of the output level is a monotonically decreasing function of the signal level with a mean level equal to the gain. This produces an additional variance in the signal which increases the noise by a factor of $\sqrt{2}$. We would normally expect that the variance in a measurement containing N photons is $\sqrt{N}$. With the EMCCD we find that this variance is $\sqrt{2N}$. The background to this process is described by Basden et al [2] who have shown that this is equivalent to reducing the quantum efficiency of the CCD by a factor of 2. However, provided we are working with images with a mean signal level much less than one photon detected per pixel per frame then each photon may be identified unambiguously. By replacing its value with a constant equal to the gain we can eliminate the variance due to the multiplication process. That effectively restores the quantum efficiency to its normal level. In fact it is not quite as simple as that. We need to apply a threshold above which each event is accepted as a photon and below which any signals are assumed to be spurious. As the most probable value of gain in the multiplication register is unity this thresholding inevitably means that we will be discarding some genuine events which have experienced a relatively low amplification. In practice this means that with a typical gain level of 1000-2000 times working in photon counting mode then the effective quantum efficiency achieved is 85-90% of the true quantum efficiency of the EMCCD.

## 3. CLOCK INDUCED CHARGE

Even in total darkness a dark signal is generated within a CCD because of its intrinsic temperature. With deep cooling, for example between -100°C and -140°C, and by operating the device in inverted mode, the dark current may be reduced to a very low level indeed (less than one electron per pixel per hour). The only downside of cooling in this range with E2V Technologies CCDs is a slight loss of quantum efficiency at wavelengths longer than ~800 nm. There is another source of unwanted signal within a CCD that is normally invisible when a device is operated with relatively long exposure times and infrequent readout. The electronic charge is transferred across the CCD by toggling the voltage on groups of electrodes. Fast readout rates imply fast rise and fall edges on the clocks and it becomes very difficult to terminate the driven clock lines at the CCD properly. Ringing will inevitably occur and the momentarily higher clock levels because of these edges can dislodge electrons from within the CCD itself. Careful driver design together with minimising the amplitude of the clock swing can minimise this spurious charge generation. It is very sensitive to the amplitude of the parallel clock swings particularly. Increasing the parallel swing from 12.5 V to 16.5 V can increase clock induced charge by an order of magnitude. However, when running one of these devices in photon counting mode and at high gain, the peak signal level in the image and store regions of the CCD will inevitably be extremely low indeed, typically one photon per pixel maximum. This means that it is possible to use significantly smaller parallel clock swings that would normally be inadequate to transfer the full well potential of the device. Typically a clock swing of 10-11 V will give excellent charge transfer efficiency while minimising clock induced charge. Indeed our experiments in Cambridge suggest that parallel region clock induced charge can be reduced to levels where it is almost negligible, considerably less than one event per pixel per thousand frames.

In the same way, the serial register can also generate clock induced charge. In this case the induced charge is generated on average halfway down the multiplication register. This means that the gain that such induced electrons are subjected to is, on average, the square root of the gain of the entire multiplication register. This gives clock induced charge events generated in the multiplication register rather different characteristics to those of the electrons that have passed along the entire length of the register. These events are therefore an average significantly lower and the thresholding can be effective at reducing the contributions substantially. With care, therefore, it is possible to reduce clock induced charge levels even at high pixel rates to levels below 1 electron per pixel per thousand frames.

We also find that even for relatively low speed operation (5-10 MHz pixel rate) great care has to be taken to minimise the length of the wires or printed circuit board tracks connecting the drivers to the device. We find that with careful design it is possible to operate well above the manufacturers specified maximum pixel rate (for the CCD 201 this is 13-

20 MHz). We routinely operate our CCD201s at 30 MHz. At that frequency the gain of the output amplifier is beginning to tail off but the charge transfer efficiency is still excellent at these frequencies so it is perfectly straightforward to use that readout rate routinely. Generally clock induced charge can be minimised by careful control of clock edges while at the same time arranging that the clock high periods are kept to a minimum. It appears that the carriers take some time to be encouraged out of the lattice so fast clocks improve CIC unless they are also accompanied by excessive clock pulse over-shoot.

## 4. EMCCD CAMERA DESIGN.

Some electronic design engineers, familiar with slowscan CCD camera design may feel intimidated by the challenges of high-speed design particularly when there is a requirement for high-voltage clocking of one of the electrodes, an essential part of an EMCCD camera. Fortunately there is a great deal of high quality support information available from device manufacturers and component manufacturers. Modern digital SLR and video cameras are built to very high standards and the integrated circuit parts available for these cameras are often remarkably capable. Many of these components have considerable potential for use in high-quality astronomical cameras. This makes electronic design today much less terrifying than it has been in the past! Nevertheless it is really important not to think that it is simply a matter of speeding up a low speed design. The entire philosophy of building the camera needs to be reconsidered. For example, with slowscan cameras it is often the case that the serial transfer only occupies a small fraction of the total pixel period. At high speed all the clocks need to run continually. Integrated circuit analog front-end components such as the Analog Devices AD9824 which is used in the Cambridge EMCCD camera are very capable but are targeted at a specific readout pixel rate so that the benefits usually achieved by running more slowly can be relatively modest.

With high-speed operation the standard astronomers approach of building the readout electronics in a separate box and connecting that to the camera head by a nice long cable is asking for trouble. A 30 MHz pixel rate means that clock rise and fall times have to be at most a few nanoseconds in extent so great care is needed to place the driver electronics as close to the head as possible and connect it with a well matched conductor whose position cannot change (and remember that with Teflon covered wires inside a vacuum dewar, disassembling and reassembling the head may well change both positions and therefore the capacitance of that wire to ground and to other wires. This can significantly affect the exact form of the clocking waveform that reaches the device). In Cambridge we prefer now to use a printed circuit board onto which the EMCCD is mounted with tracks on internal layers which allow connection to the driver electronics cards that sit outside the dewar (figure 2). This allows a highly repeatable driver set up. A further complication of running at high frequency with these extremely weak signals (the E2V CCD201 has an output sensitivity of 1.4 µV per electron, and probably closer to 1 µV per electron at 30 MHz) is the fact that it is very difficult to examine the CCD output waveform even with a high-quality oscilloscope.

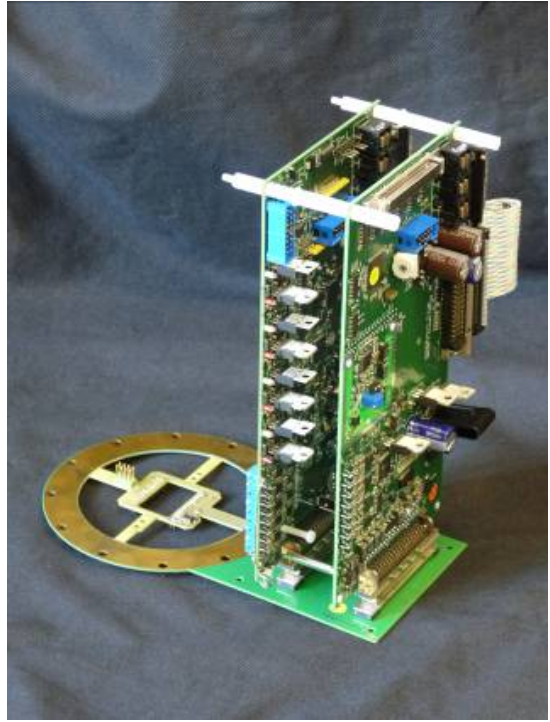

Figure 2: The internal structure of the EMCCD cameras built in Cambridge. The EMCCD itself is mounted on a printed circuit board which extends through the wall of the liquid nitrogen vacuum dewar with all the electrical tracks on internal PCB layers. The multilayer PCBs (one for the ARM microprocessor, sequencer and clock drivers, plus one for the analogue signal processing plus bias and monitoring supplies) are connected through 68 way edge connectors on the PCB.

The high-voltage clock design also represents a number of significant challenges. The most obvious way to drive it is with a relatively straightforward switching circuit but, at higher frequencies, the power dissipated by such a circuit becomes unacceptably high. An alternate approach is to use a stable sine wave generator that is synchronised and phased with the other clocks. In Cambridge our original work used switching circuitry, but the power demands at 30 MHz became unacceptable. The need to operate cameras at altitude where the air is thinner and cooling somewhat more challenging forced us to use sine wave drivers. In fact, we now find that they are preferable at almost any frequency and give good, reliable and consistent performance. Care has to be given also to the choice of the ground voltage level since, of course, the range of voltages used are completely at the disposal of the engineer. The CCD chip itself does not have any ground pads or any other structure such as the device package that has to be connected to ground other than through the external pins. The voltages that need to be produced by the power supply and regulated to provide the high-voltage clock levels can produce safety issues (the European Low Voltage Directive[3] has safety requirements for AC voltages above 50 V and DC voltages above 75 V on the inputs or outputs from a unit). With differential voltages on the printed circuit board that can be in excess of 50 V, care has to be taken to ensure that tracks are well enough separated so there is no risk of arcing between tracks particularly at the lower atmospheric pressure found at high altitude observatories.

The stability of the high-voltage clock high level is important. Changing the high-voltage clock high level by 0.3 V at high gain (~2000) can double the device gain. In any photometric application gain calibration is important. In addition the gain depends on the temperature of the CCD itself. Although temperature control may be provided the effect of clocking a CCD at high speed injects energy into the device which heats it up. In practice, therefore, it is important to have leave the CCD running continually so that it is not subjected to the temperature changes that would be caused by starting and stopping the device. If the EMCCD camera is to be used in analogue (non-photon-counting) mode then the stability of the high-voltage clock level is a very important indeed. However if it is being used in photon counting mode then these tolerances may be relaxed as the gain simply needs to be high enough to make the identification of a single photon event unambiguous. At event will be replaced by some mean number and, provided the gain is not changing very much, the thresholding necessary will not have to be adjusted greatly. Indeed it is conceivable that this threshold level could be computed for each and every frame dynamically further minimising the effects of any gain variations.

Commercial EMCCD cameras are available from a number of manufacturers. They are usually important that they work satisfactorily at unity gain as well as high gain so must deliver the standard EMCCD full well images. Inevitably the clock voltages needed for this are larger than they absolutely need to be if one is working in photon counting mode. This means that there is a greater risk that clock induced charge will be significant. There is also the difficulty of customising the operation of the camera, in particular its pixel rate and/or clocking bias levels, and so increasingly astronomers will turn to their design engineers to deliver EMCCD hardware that is tailored specifically to their own requirements as they have done so often in the past with conventional slowscan CCD camera systems.

## 5. EMCCD APPLICATION: LUCKY IMAGING COMBINED WITH ADAPTIVE OPTICS

From any photon counting application a high frame rate is desirable. A low frame rate increases the risk that adjacent pixels will be filled by another photon or that individual pixels will have more than one electron within them. Small area EMCCDs such as the CCD60 are able to run at kilohertz frame rates but the CCD201 with 1024 x 1024 pixels even at 30 MHz will only produce a 25 Hz frame rate. Until large area devices with multiple parallel readout registers are available, the only way of building a camera with a large contiguous field of view is by optically butting them in a way pioneered for the original WF/PC instrument on the Hubble Space Telescope. This is relatively straightforward although it does require a somewhat more complicated optical configuration. Multiple detectors may be mounted within a single dewar.

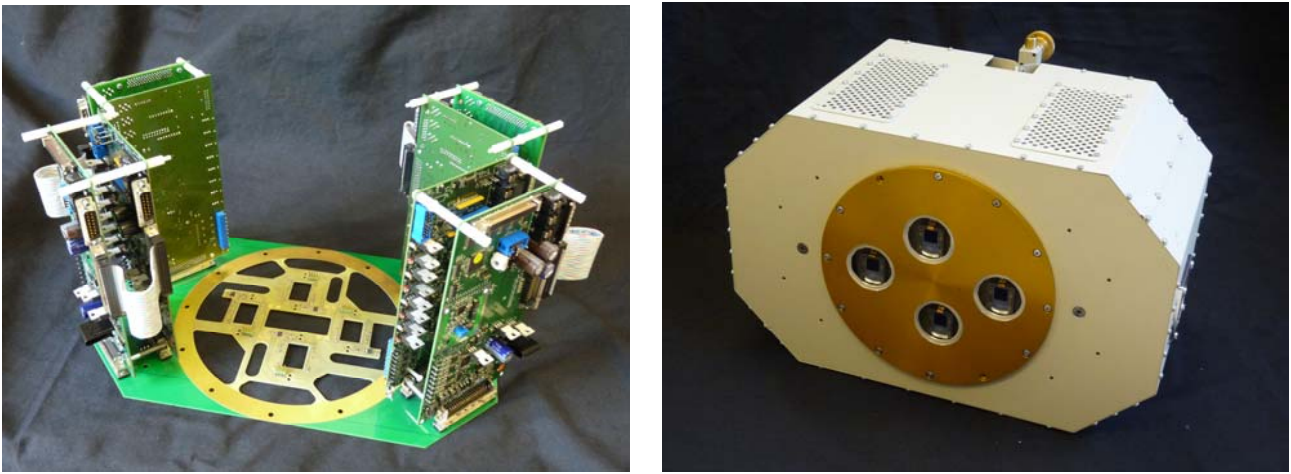

Figure 3: The quad EMCCD camera also uses a printed circuit board that extends through the wall of the vacuum dewar so that the four sets of driver electronics may be mounted as close to the EMCCDs as possible.

We have taken this approach in Cambridge to developing a wide field rectangular format (4096 x 1024) camera used on the Nordic Optical Telescope on La Palma and for a wide field 2048 x 2048 camera as part of the AOLI project for the William Herschel 4.2 m and GTC 10.5 meter telescopes, also on La Palma[9]. The requirement that the driver electronics should be as close to the EMCCDs as possible makes the use of a custom camera design essential (figure 3).

Lucky Imaging was first described by Hufnagel[4] in 1966 and given its name by Fried[5] in 1978. It describes the technique of taking images fast enough to freeze the motion caused by atmospheric turbulence and then selecting the sharpest images to synthesise a composite image much sharper than would be obtained with conventional long exposures. High-speed photon counting cameras based on EMCCDs have only recently made this technique viable. A number of groups have built lucky imaging cameras, all using EMCCDs, and have published a large number of scientific papers on the results. There are a number of other papers concerning lucky imaging being presented this week at the SPIE meeting in Amsterdam (2012). On a 2.5 m telescope, the same size as the Hubble Space Telescope, ground-based telescopes using lucky imaging selection are able routinely to equal or exceed the resolution of the HST in the visible[6]. Other high-resolution imaging techniques such as adaptive optics have not been able to achieve this despite the very substantial amount of money expended on the technique. An example is shown in figure 4.

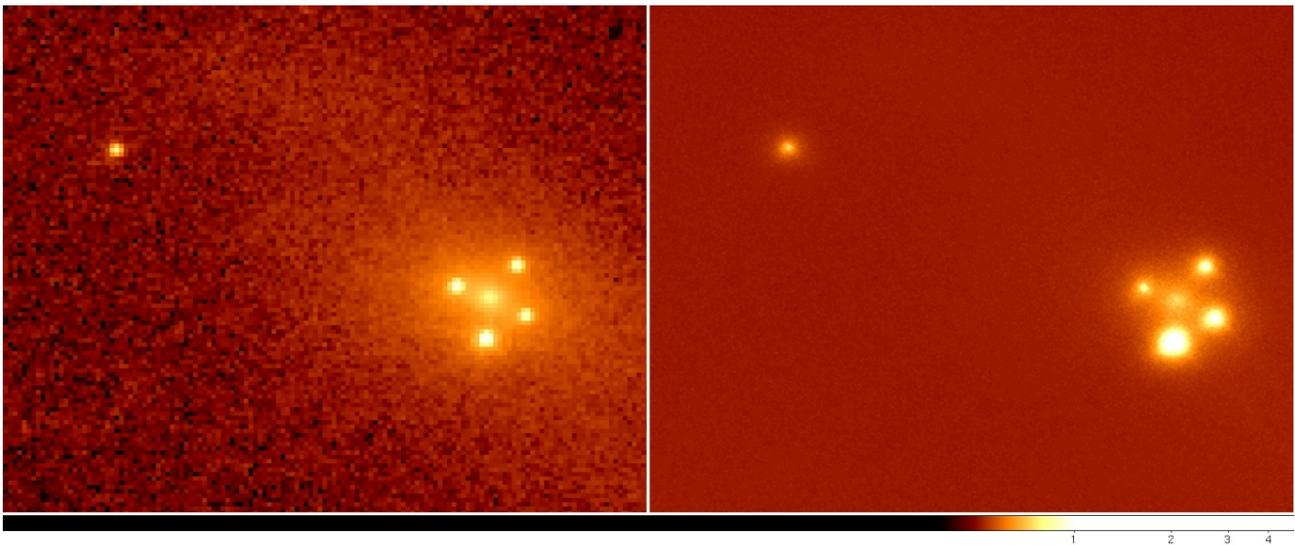

Figure 4: Images of the Einstein cross, showing four gravitationally lensed images of a distant quasar magnified by the core of a relatively nearby bright galaxy. The left-hand image was taken with the HST Advanced Camera For Surveys and the right hand image taken with the Cambridge LuckyCam attached to the NOT 2.5m telescope on La Palma.

The big challenge now is to extend these techniques to larger telescopes and higher resolution. Unfortunately, while the probability of a sharp image is acceptable on a 2.5 m telescope (typically 3-30% of the images can be used under reasonable conditions), the probability of an equivalent a sharp image on a significantly larger telescope is so small as to be quite hopeless. This is because there are too many turbulent cells across the diameter of the telescope and the chance of a substantial number sorting themselves out at the same time becomes negligible. With a turbulent atmosphere most of the power is in the largest scales (tip-tilt, defocussing, astigmatism, coma etc). Removing any of these low order terms effectively increases the diameter of a typical cell, defined as the mean size of a patch of the wavefront entering the telescope with a phase variance of one radian squared. As each order is removed, for example by a low order adaptive optic system, the effective cell size increases. Eventually the number of corrected cells across the diameter of the telescope is reduced to a level comparable to that on a 2.5 m telescope and Lucky Imaging may be used again efficiently.

In order to try these techniques we attached our Lucky Imaging camera to the Palomar 5 m telescope behind their PALMAO instrument, a low order high-speed adaptive optics system. The tests were very successful[12], producing the highest resolution image ever taken of faint targets in the visible with any telescope on the ground or in space (figure5).

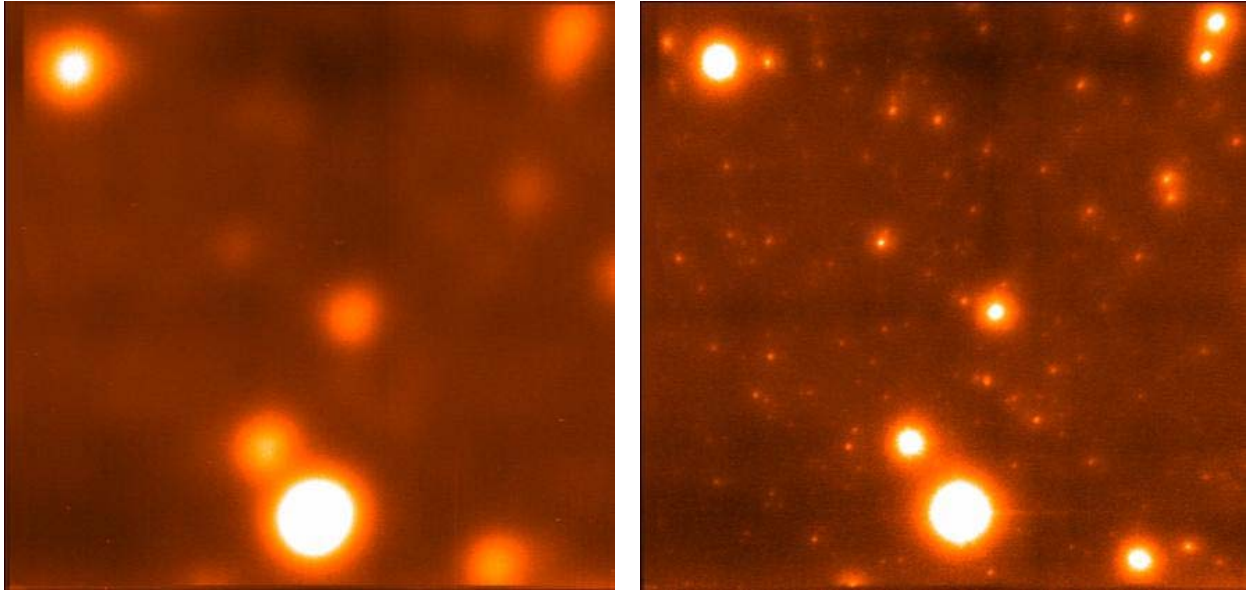

Figure 5: The Globular cluster M13 as imaged conventionally by the Palomar 200 inch telescope (left), followed by M13 as imaged with the Lucky Camera behind an adaptive optics system on the Palomar 200 inch telescope (right). The natural seeing was about 0.65 arcsec, and the Lucky/AO image has a resolution of about 35 milliarcsecs or about three times that of the Hubble Space Telescope. This image is the highest resolution image ever taken of faint targets with any optical or infrared telescope on the ground or in space.

The resolution obtained was about 3 times that of the Hubble Space Telescope at 35 milliarcseconds with natural seeing of about 650 milliarcseconds, a dramatic improvement in image quality. The problem with the lucky camera plus PALMAO approach is that the AO system which uses a Shack-Hartmann wavefront sensor requires a relatively bright and therefore relatively scarce reference star that would limit the application of the technique to a very small fraction of the sky area[10]. It was clear that a radically different approach was needed if we were to be able to use faint reference stars at a level where we could be fairly confident of finding one in our field of view even at high galactic latitudes. Studies by Racine[7] have shown that curvature sensors are much more sensitive than Shack-Hartmann sensors particularly for low order correction. Curvature sensors work by recording the light from the reference star on either side of a pupil plane. A patch that brightens as it goes through the pupil corresponds to a converging wavefront whereas if it darkens as it goes through the pupil it corresponds to a divergent wavefront. Combining all the patches across an aperture allows the curvature to be determined in a way that corresponds exactly to the adjustments needed by the deformable correction mirror. The propagation of the wavefront through the pupil is, however, non-linear so that it helps greatly if two additional image planes are used further from the pupil plane.

The four near-pupil plane images are generated by splitting the light four ways and recording them on separate EMCCDs. The typical pupil plane image size is less than 256 x 256 pixels so that it is possible to record these images at faster than 100 frames per second. A wavefront is fitted to these data and the result used to drive a deformable mirror in the optical train. The advantage of a curvature sensor comes from the fact that at the lowest light level it is still possible to derive approximate wavefront error data over fairly large scales across the aperture of the telescope while a Shack-Hartmann sensor reaches its minimum sensitivity limit at a much higher light level because it senses the wavefront over much smaller cell sizes. This is particularly true for low order adaptive optics correction which is what we wish to achieve since nearly all the turbulent power in atmospheric turbulence is concentrated on the largest scales[11]. Our simulations[8,9] suggest that a limiting magnitude for the reference star of I ~ 17.5-18 on the WHT 4.2 m telescope and I ~ 18.5-19 on the GTC 10 meter telescope will be achievable. The curvature sensor design is substantially achromatic particularly for low order corrections permitting a wide bandpass to be used for the curvature sensor. At these magnitude levels it is possible to find a reference star relatively easily in most of the fields, a key requirement for an instrument to be of use over the whole sky.

An instrument[9] called AOLI (Adaptive Optics assisted Lucky Imager) using four EMCCDs has a 2048 x 2048 science camera with a field of view of up to 2 arc minute square and a dual EMCCD curvature sensor where two of the near-pupil images are projected onto each detector is presently being built for use on the WHT and GTC. AOLI is designed to produce images with an angular resolution of ~40 milliarcseconds and ~15 milliarcseconds respectively in I band.

## 6. CONCLUSIONS.

The development of electron multiplying CCDs is beginning to open up a variety of opportunities for high time resolution imaging and spectroscopy on the latest generations of optical telescopes. Their capacity to work in photon counting mode without compromising greatly the high quantum efficiency and excellent charge transfer capability of CCDs is enabling scientific studies that have not hitherto been possible. The challenges involved in designing cameras to work with EMCCDs at the lowest light levels are entirely manageable. EMCCDs have remarkable potential for opening up new fields of astronomical research in a number of important areas.

## ACKNOWLEDGEMENTS

It is a pleasure to acknowledge the help of staff of the Nordic Optical Telescope on La Palma over a number of years, and also to acknowledge the help the staff of the Palomar Observatory in California.